# Designing Social VR: A Collection of Design Choices Across Commercial and Research Applications

RYAN HANDLEY, Oakland University, USA

BERT GUERRA, Oakland University, USA

RUKKMINI GOLI, Oakland University, USA

DOUGLAS ZYTKO, Oakland University, USA, zytko@oakland.edu

Social VR has experienced tremendous growth in the commercial space recently as an emerging technology for rich interactions themed around leisure, work, and relationship building. As a result, the state of social VR application design has become rapidly obfuscated, which complicates identification of design trends and uncommon features that could inform future design, and hinders inclusion of new voices in this design space. To help address this problem, we present a taxonomy of social VR application design choices as informed by 44 commercial and prototypical applications. Our taxonomy was informed by multiple discovery strategies including literature review, search of VR-themed subreddits, and autobiographical landscape research. The taxonomy elucidates three design areas—the self, interaction, and the environment—and organizes 10 feature categories and 38 feature variations. The paper reflects on how the taxonomy can support involvement of new designers, and considers possible expansion and use by researchers.

## 1 INTRODUCTION

Social virtual reality (VR) refers to 3D panoramic virtual spaces, typically accessed through head-mounted displays, that are primarily intended to support social interaction. Despite being discussed in the literature over decades [7,8,16,17], only in the past few years have social VR applications become widely accessible to the public due to affordable VR equipment such as the Oculus Rift and HTC Vive [55]. Research into public adoption of social VR has confirmed its potential for augmenting social interaction across a variety of user demographics [3,45,52,76]. Yet emerging obstacles in the way of user satisfaction and wellbeing have also been discovered, reminding us that social VR application design is still in need of improvement. Identity-based harassment [11,20,60] and privacy risks [48], particularly regarding marginalized demographics, are becoming prominent issues. Other motivations for design include barriers to participation by older adults [3] and limitations to social activities [45] and personal expression [52].

Social VR research has recognized that a barrier to designing solutions for emerging issues, and progressing social VR more broadly, is a lack of clarity regarding the state of social VR application design [50,51]. Questions like "how have social VR applications been designed?" are difficult to answer because commercial applications are relatively new, and numerous research prototypes are scattered across the literature. This creates a barrier for new voices in design, especially from demographics at risk of emerging issues like older adults and children, and can obfuscate the full range of design possibilities even to seasoned designers.

Following early efforts in the 1990s [7,16], documentation of social VR design choices has been reinvigorated in the last few years [36,50,51,66]. The latest documentation encompasses recent commercially available applications and with a focus on particular types of design elements such as non-verbal communication features [66] and design choices deemed supportive of pro-social interaction [50]. While these efforts lend much needed coherence to the design space of relatively new social VR offerings, non-commercial social VR applications developed for research purposes or private use over the past three decades are absent.

This paper extends efforts into mapping social VR design by presenting a taxonomy of social VR design choices informed by 44 commercial and non-commercial applications. We use the word taxonomy quite deliberately to emphasize the importance of succinctly categorizing design choices to support rapid comprehension and review for informing future designs.

## 2 BACKGROUND

### 2.1 Making VR Social

The fundamental driver behind advancements in virtual reality (VR) has been the capacity for presence, or the feeling of "being there" [30,43]. Introducing multiple users into the same VR environment for interaction extends this potential to include social presence or copresence; the feeling of being present and interconnected with other people [10,38]. "Social VR" [50] is the modern terminology for applications that enable interpersonal contact in 3D panoramic virtual spaces, typically through head-mounted displays and other technologies to capture body movement and communication. The term has come to refer more specifically to VR environments designed primarily for supporting social interaction [2,20,58]. This differentiates from solo VR experiences [1,44] and VR games [18], the latter in which interaction is sometimes possible but not necessarily the main motivation for design or use.

Efforts to design social VR applications have followed two tracks: applications designed for research purposes or private use, and commercial applications that are available to the public. The research track has a longer history, gaining popularity in the early to mid 1990s [6,19]. Such applications were originally called collaborative virtual environments (CVEs) [4,7,8,12,15,17] and originated from academic institutions [7], although some early industry applications also existed [39]. While the public was largely unable to enjoy social VR in the 1990s due to barriers in technological capabilities and cost, early prototypes like DIVE and MASSIVE [6,7,16] were responsible for foundational advancements in user embodiment and expression that are still present in social VR applications today. Design and research with prototypical social VR applications continued into the 2000s and into the modern day [2,13,32,62,63,67,73].

The release of affordable VR headsets such as the Oculus Rift and HTC Vive in 2012 and 2016 paved the way for commercial social VR applications [55]. Popular examples include AltSpaceVR, Rec Room, and VRChat, among others. Relative to earlier CVEs that intended to support specific contexts of interaction like collaborative work [16], interactions in today's commercial social VR applications can be broad and, at times, unpredictable. They typically produce random interactions amongst strangers through public virtual areas that support a massive number of simultaneous users. Some commercial social VR applications also support interaction with friends, family, and preexisting contacts through private and public social areas.

The advent of commercially available social VR applications has enabled empirical research into how the public use, and want to use, social VR. Some of this research has affirmed social VR's capacity for social presence and potential to augment social life [20,48,52], and for a variety of demographics including children [46], older adults [3], long distance romantic couples [76], and marginalized identities [20]. Discoveries about early public adoption of social VR have not all been positive however. Reports of online sexual violence and harassment have propagated through popular media and research [5,54,60], particularly effecting women and ethnic minorities [20] and taking various forms like verbal harassment, simulated touching, and displaying graphic content in the virtual environment [11]. Furthermore, Maloney and colleagues found an alarming lack of barriers around child-adult interactions [47], which can render children susceptible to sexual predators and inadvertent exposure to adult content. Beyond interpersonal harm, empirical research has conveyed other user struggles and preferences for design progress. Movement mapping techniques [65] can invoke ageing stereotypes and pose a barrier of use for older adults [3]. Limitations with non-verbal communication have posed difficulties during one-to-one interactions between friends [52]. Users have also asked for general increases in connectivity in and out of social VR [45].

## 2.2 Documenting Social VR Design

Since at least the 1990s social VR researchers have recognized the value of documenting the social VR design space as to way to inform design progress and solutions to emerging user problems. Benford and colleagues unpacked the designs of the DIVE and MASSIVE prototypes in 1995 as to a way to inform and motivate pursuit of design issues around user representation and expression in social VR [7]. Churchill and Snowdon also documented DIVE and MASSIVE, among other leading CVEs of the day, in their 1998 review of the design space [16].

In the last three years researchers have reinvigorated pursuits to document social VR design, with a particular focus on new, commercially available applications [36,50,51,66]. McVeigh-Schultz and colleagues' mapping of "the social VR design ecology" in 2018 and 2019 was partially motivated by issues of interpersonal harm in social VR, noting that social VR design plays a role in "shaping particular kinds of social practice" [51]. Informed by autobiographical landscape research (the researchers' personal use of social VR applications) [51] and interviews with social VR industry experts [50], they discovered a series of design choices in seven commercially available social VR applications deemed supportive of pro-social interaction. These design choices broadly involved "aesthetics of place" to make users feel like they were situated in familiar environments, "communicative affordances and social mechanics" like emojis and handshake gestures, and features intended to shape social norms and ease users into socializing such as grilling virtual burgers over a campfire. Kolesnichenko and colleagues also used that industry expert interview study to more pointedly document the design of user avatars in commercial social VR, including the ability to create humanoid and non-humanoid avatars [36]. Tanenbaum and colleagues, noting limitations around user expressiveness in virtual environments, developed an inventory of non-verbal communication design choices in 10 commercial social VR applications [66]. These choices were organized into categories including movement & proxemic spacing, facial control, and gesture & posture.

In this paper we extend recent efforts to catalogue the social VR design space in two ways. One, we expand documentation of social VR design to include both commercial applications and non-commercial applications designed for research or private purposes. Commercial social VR applications are less than 10 years old, yet non-commercial social VR applications stretch back over three decades. Second, our efforts are purposely devoid of intent to find design choices "for" a particular purpose or end goal, be it pro-social interaction [50], user embodiment [7], or non-verbal communication [66]. Of course, a benefit of these narrow lenses applied in prior work is specificity and granularity in discovered design choices. However, design choices outside of those lenses can be left undiscovered. We intended to not only document design choices across commercial and prototypical social VR applications, but to categorize them into rapidly digestible design areas—we call this a design taxonomy.

## 3 METHOD

### 3.1 Relation to Prior Work

This paper represents the continuation of a previously published work [33]. This paper offers an expanded literature review, a review of Reddit, and autobiographical landscape research.

### 3.2 Summary of Method

We conducted a four-step discovery process over a 12-month period to identify social VR applications and design choices to serve as the basis of a design taxonomy. These steps were: 1) a search of scholarly databases to identify literature about social VR applications; 2) a review of reference lists from the literature discovered in the previous step to identify further literature describing social VR application designs; 3) a review of VR-related online communities on Reddit (called subreddits) to identify additional social VR applications and features; and 4) autobiographical landscape research [19] with commercially available social VR applications.

A total of 44 social VR applications were discovered (see tables 1 and 2). Fifteen were identified as publicly or commercially available applications. The other 26 were non-commercial applications created for research or private use. A total of 38 unique design choices within the social VR applications were discovered. A preliminary taxonomy of social VR design choices was created after step 1 (scholarly literature review) through a card sorting exercise [64]. This involved the research team collaboratively organizing and re-organizing discovered design choices into emerging themes based on their commonalities. The taxonomy was iteratively refined and expanded through further card sorting sessions after each subsequent step. We elaborate on each discovery step below.

### 3.3  Search of Scholarly Databases for Literature on Social VR Application Design

We conducted a literature review using the ACM Digital Library, IEEE Xplore Digital Library, Google Scholar, and our university library's online database (comprising over 970,000 sources of journals, books, and proceedings). The following search terms were used on each site: *social VR, social virtual reality, virtual reality, virtual worlds, Oculus Rift, HTC Vive, and Collaborative Virtual Environment*. Five researchers individually reviewed 10 pages of search results for each term per website, resulting in a total of 4550 search results reviewed per person. Discovered literature was saved for a full review if the title or abstract mentioned 1) the term "social VR" and/or 2) a VR environment seemingly designed, or studied in its capacity, to facilitate interaction between users in VR. After removing redundant discoveries this resulted in a corpus of 39 publications that were fully reviewed to identify any VR applications and features of those applications intended to support interaction between users. A total of 29 applications were discovered from this review. Nine were identified as publicly or commercially available applications in their respective literature. The other 20 were research prototypes intended for private use.

### 3.4  Review of Reference Lists

The reference lists of the 39 publications discovered in step 1 were reviewed to identify additional literature. Two researchers collectively selected referenced literature for review if the title of the reference or content connected to its in-text citation suggested that the source may contain information about VR environments designed or studied in their capacity to facilitate social interaction. In total, 78 cited texts were selected for full review. Eleven additional social VR applications were discovered; 2 commercially available applications and 9 prototypical or otherwise non-publicly available applications. This review also reinforced and expanded the team's understanding of previously discovered social VR applications and design choices.

### 3.5  Review of Reddit

Reddit's API was used to collect posts and associated metadata from several online communities (subreddits) related to social VR. A set of 13 public subreddits related to VR was compiled by using Reddit's native subreddit search feature to search for subreddits based on the terms *social VR, social virtual reality, virtual reality, virtual worlds, Oculus Rift, HTC Vive*, and *Collaborative Virtual Environment*. Two researchers familiar with Reddit then searched for the term *social* on each of the 13 subreddits, sorted the results by "Top posts of all time," and inspected the title and content of the first 50 posts to identify social VR applications or design choices (a total of 650 posts were inspected per researcher). Through this process 4 additional social VR applications were discovered, all of which were commercially available. Information that reinforced understanding of previously discovered social VR applications was also collected.

Table 1. A list of social VR applications that informed the taxonomy.

| 1 | **Mozilla VR:** browser-based social VR platform [34,50] | Commercial |
|---|---|---|
| 2 | **Anyland:** platform informed by communication between developers and users [50] | Commercial |
| 3 | **High Fidelity:** encourages users to build their own virtual worls [31,50–52] | Commercial |
| 4 | **VRChat:** known for harassment and unpredictable social encounters [50,51] | Commercial |
| 5 | **AltSpaceVR:** accessible on a variety of VR devices [31,50–53,61] | Commercial |
| 6 | **Rec Room:** modeled after physical world recreation centers [50,51,53] | Commercial |
| 7 | **Facebook Spaces:** Private space for personal contacts to interact [31,40,50,51] | Commercial |
| 8 | **vTime:** Open rooms in which users can converse around a table [52,60] | Commercial |
| 9 | **Oculus Rooms:** Space for interaction with an apartment-like feel [52] | Commercial |
| 10 | **NeosVR:** Private space for friends to interact [69] | Commercial |
| 11 | **Sports Bar VR:** Space for interaction with a sports bar-like feel [72] | Commercial |
| 12 | **Sansar:** Open world with interaction between roaming users [70] | Commercial |
| 13 | **Half + Half:** Open world that facilitates social interaction with activities [71] | Commercial |
| 14 | **dVS:** 1990s-era commercial social VR from DIVISION [7,26] | Commercial |
| 15 | **Collaborative Workspace:** social VR to support cooperative work [7,68] | Commercial |
| _16_ | **Training in IVR:** social VR environment for virtual training [23] | **_Non-commercial_** |
| _17_ | **Wooded Manniquin:** social VR environment to test effects of avatar realism [37] | **_Non-commercial_** |
| _18_ | **VR Classroom:** For student learning and socialization [41] | **_Non-commercial_** |
| _19_ | **Social MatchUP:** For users with Neuro-developmental Disorder (NDD) [42] | **_Non-commercial_** |
| _20_ | **Faceless Wooden Mannequins:** assessed effect of nonverbal cue absence [57] | **_Non-commercial_** |
| _21_ | **Immersive Deck:** social VR for team building exercises [75] | **_Non-commercial_** |
| _22_ | **Trust Test:** social VR environment to test trust between avatars [22] | **_Non-commercial_** |
| _23_ | **TogetherVR:** web-based social VR platform with photorealistic avatars [27–29,56] | **_Non-commercial_** |
| _24_ | **PrototypingVR:** collaborative virtual design/prototyping [35] | **_Non-commercial_** |
| _25_ | **HOLO-DOODLE:** painting with others in VR [49] | **_Non-commercial_** |
| _26_ | **DYNECOM VR:** Users communicate with visualized brainwaves [58] | **_Non-commercial_** |
| _27_ | **Embodied VR:** collaborative apartment planning and furniture layout [63] | **_Non-commercial_** |
| _28_ | **Virtual Dancing:** reconstruction of physical-world space in VR [65] | **_Non-commercial_** |
| _29_ | **The CAVE:** social VR environment that combines projection with shutter glasses [17] | **_Non-commercial_** |
| _30_ | **ImmersaDesk:** built on CAVE library software [17] | **_Non-commercial_** |
| _31_ | **TheDomeCityMOO:** Participants assume different traits and roles in a VR city [17] | **_Non-commercial_** |
| _32_ | **Holojam in Wonderland:** social interaction in the context of a theater performance [24] | **_Non-commercial_** |
| _33_ | **Active Worlds:** "Open world" concept for users to exhibit motiond [4] | **_Non-commercial_** |
| _34_ | **OnLive Traveler:** Head-only avatars communicate via facial expressions, voice, text [4] | **_Non-commercial_** |
| _35_ | **Sync VR:** Social VR application for analyzing synchronization and social connection [67] | **_Non-commercial_** |
| _36_ | **MASSIVE-1:** One of the first 1990s-era social VR environments [7,14,15,25] | **_Non-commercial_** |
| _37_ | **MASSIVE-2:** Upgraded version of MASSIVE-1 with additional functionality [8,9,16] | **_Non-commercial_** |
| _38_ | **DIVE:** Modular social VR platform [7,21] | **_Non-commercial_** |
| _39_ | **NPSNET:** Human interaction in the context of collaborative military activities [16,77] | **_Non-commercial_** |
| _40_ | **Diamond Park:** Social interaction themed around virtual bicycling in a park [16,74] | **_Non-commercial_** |
| _41_ | **Tele-Immersive VisualEyes:** created by General Motors for collaborative CAD [39] | **_Non-commercial_** |
| _42_ | **Virtual Temporal Bone:** physicians to teach and interact with medical students [39] | **_Non-commercial_** |
| _43_ | **CAVE6D:** users can "jointly visualize, discuss and interact with datasets" [39] | **_Non-commercial_** |
| _44_ | **Humanoid Avatars:** a DIVE derivative to test effects of avatar realism [21] | **_Non-commercial_** |

.

## 3.4 Autobiographical Landscape Research

Autobiographical landscape research involves documentation of self-usage as a research tool for exploring design choices [51]. In our case this involved two researchers documenting their use of previously discovered social VR applications accessible on the Oculus Rift. The applications used were: AltspaceVR, Anyland, Facebook Spaces, High Fidelity, NeosVR, RecRoom, Sansar, VRChat, and Vtime. An early version of the taxonomy was used to build a task-based checklist to examine design choices in each application. This checklist had three sections: user representation, interaction, and the surrounding environment. Exploration of each social VR application involved two researchers: one using the application with an Oculus Rift headset ("the user"), and the other guiding the user through the task-based checklist while taking field notes (the social VR environment was viewable to this researcher through a large external monitor). Review of the recordings and field notes led to identification of four new features and substantial refinement to how the taxonomy was organized.

## 4 TAXONOMY OF SOCIAL VR DESIGN

The social VR design taxonomy is organized into three design areas: 1) *the self* – how users are represented within the virtual environment; 2) *interaction* – how users can communicate or exchange information; and 3) *the environment* – the virtual surroundings in which users are located. Each table is organized into three columns. The first column (left) represents the primary design area. The second column (middle) organizes feature categories pertinent to the =design area, and the third column (right) organizes individual design choices or variations within each category. Numbers next to each feature variation indicate the social VR applications that were discovered to include the design choices. The absence of an application number does not necessarily mean the application is devoid of that feature variation, but rather that its existence within was not identified in our discovery efforts. Bold/underlined/italicized numbers refer to non-commercial applications.

Table 2. The section of the social VR design taxonomy pertaining to "the self" – how users are represented. Refer to Table 1 to connect social VR application numbers with their names.

| Design Area | Features | Variations of Features |
|---|---|---|
| The Self | Avatar Representation | Partial Body Avatars 1, 5, 6, 7, 9, 11, 14, *__32, 34, 43__* |
| | | Full Body Avatars 3, 4, 5, 8, 10, 12, 13, 15, *__16, 17, 20, 22, 23, 25, 26, 27, 28, 30, 32, 35, 36, 37, 38, 40, 44__* |
| | | No Avatar *__18, 19, 24__* |
| | Avatar Customization | Preset Avatars 1, 3, 4, 5, 6, 7, 12, 14, 15, *__17, 20, 23, 26, 28, 30, 35, 37, 38, 41, 44__* |
| | | Appearance Customization 2, 3, 4, 5, 6, 7, 8, 9, 11, 12, *__36, 38__* |
| | | No Customization 13, *__16, 18, 19, 24-26__* |
| | Avatar Manipulation | Full Body Tracking 3, 4, 5, 7, 14, 15, *__16, 20, 23, 25, 27, 28, 32, 35, 37, 38, 41, 44__* |
| | | Hand-Only/Minimal Tracking 1, 2, 3, 4, 5, 6, 7, 8, 9, 10, 11, 12, 13, 14, 15, *__16, 21, 24, 25, 27, 39, 40__* |
| | | No Tracking/Minimal Tracking *__18, 19, 26__* |
| | Avatar Locomotion | Teleporting 2, 3, 4, 5, 6, 11, 13, *__40__* |
| | | Walking 1, 2, 3, 4, 5, 6, 9, 10, 11, 12, 13, *__20, 21, 28, 36, 37, 38__* |
| | | Climbing 10, *__36__* |
| | | Flying 3, 10, 13, *__39__* |
| | | Bicycling *__40__* |
| | | No Traversing 7, 8, 15, *__19, 26, 35__* |

Table 3. The section of the social VR design taxonomy pertaining to "interaction with others" – how users can communicate or exchange information with one another. Refer to Table 1 to connect social VR application numbers with their names.

| Design Area | Features | Variations of Features |
|---|---|---|
| Interaction | Communication Privileges | Muting Other Users 2, 3, 4, 5, 6, 7, 8, 11, 12 |
| | | Blocking Other Users 1, 2, 3, 4, 5, 6, 8, 9, 12 |
| | | Space control 3, 4, 6, 11 |
| | | Adding/Deleting Other Users in Contact Lists 2, 3, 4, 5, 6, 8, 9, 10, 11, 12 |
| | | Inviting Other Users to Private Social Areas 1, 2, 3, 4, 5, 6, 7, 8, 9, 10 |
| | Communication Types | Voice 2, 3, 4, 5, 6, 7, 8, 9, 10, 11, 12, 13, 14, 15, *__16, 20, 22, 23, 24, 28, 33, 35, 36, 37, 40, 42, 44__* |
| | | Text-Based 10, 12, *__18, 33, 34, 36, 37__* |
| | | Physical Expression 1, 2, 3, 4, 5, 6, 7, 8, 9, 10, 12, 13, 14, 15, *__16, 20, 23, 25, 27, 28, 29, 30, 32, 33, 35, 37, 38__* |
| | | Visemes 2, 3, 4, 5, 6, 7, 8, 9, 10, 11, 12, 13, 15 |
| | | Transactions 3, 4, 6, 12, *__23__* |
| | | Visualized Bio-Adaptive Feedback *__26__* |
| | Activity to Scaffold Interaction | Events 2, 3, 4, 5, 12, *__19, 21, 32, 36, 37, 38__* |
| | | Recreation 1, 2, 3, 4, 5, 6, 10, 11, 13, *__25, 28, 29, 30, 31, 33, 40__* |
| | | Collaborative work/design 7, *__24, 27, 41, 42, 43__* |
| | | No Activity Scaffolding (Conversation Only) 7, 8, 9, *__16, 18, 24, 34__* |

Table 4. The section of the social VR design taxonomy pertaining to "the environment" – the virtual surroundings in which users are located and interact. Refer to Table 1 to connect social VR application numbers with their names.

| Design Area | Features | Variations of Features |
|---|---|---|
| The Environment | User Manipulation of Environment | Construct a New Virtual Space 2, 3, 4, 5, 10, 12, *__33__* |
| | | Create and Alter Objects 1, 2, 3, 4, 5, 6, 7, 9, 11, *__16, 19, 20, 21, 22, 24, 25, 27, 33, 38__* |
| | | No Environment Manipulation 7, 8, *__17, 18, 23, 26, 28, 29, 30, 31, 32, 34, 35, 36, 37__* |
| | Spawning Area | Private Area Spawning 2, 3, 4, 5, 6, 7, 8, 9, 10, *__28, 36, 37__* |
| | | Social Area Spawning 11, 12, 13, 14, 15, *__16, 18, 19, 20, 21, 22, 23, 25, 26, 32, 36, 37, 38__* |
| | | Tutorial 10, 11, 12, 13, *__40__* |
| | Openness of Environment | Public Social Areas 2, 3, 4, 5, 6, 8, 11, 12, 13 |
| | | Private Social Areas 1, 2, 3, 4, 5, 6, 7, 8, 9, 10, 13, 14, 15, *__16, 17, 18, 19, 20, 21, 22, 23, 24, 25, 26, 27, 28, 29, 30, 31, 32, 33, 34, 35, 36, 37, 38, 39, 40, 41, 42, 43, 44__* |

## 4.1 Taxonomy Section 1: The Self

This design area (table 2) encapsulates design choices enabling users to present themselves through a virtual avatar [59]. Four categories of features involving the self were discovered. *Avatar representation* refers to how much of a body, or how many body parts, a user's avatar can depict in VR. Some social VR applications depict a full body (head, arms, torso, and legs), while some depict a partial body such as only a head (see OnLive Traveler [4]) or a head with torso. Three prototypical applications provide no avatar at all, with the user's presence instead being conveyed through manipulation of objects.

*Avatar customization* refers to the user's ability to modify their avatar's appearance. Several applications provide preset avatar choices, meaning users can select from a variety of fully designed avatars. Appearance customization lets users modify individual parts of their avatar. In some cases users can upload their own content into their avatar design such as scans of their physical world appearance. For example, the prototypical application DIVE from the 1990s enabled users to upload a picture of their real face to their avatar's head [7].

It should be noted that preset and fully customizable avatars do not have to reflect a traditionally human form. In several applications, particularly those in the commercial space, avatars can depict animals, mythical creatures, robots, and even traditionally inanimate objects (see VRChat [50,51] or High Fidelity [31,50–52]). A less common design choice for avatar customization is no customization at all, meaning users are assigned a non-modifiable avatar. This was typically seen in prototypical applications.

Social VR applications also vary in *avatar manipulation*, or how users control their avatar. Some use full-body tracking cameras and hardware so that the user's physical movement is mimicked through their avatar. One unique variation comes from the prototypical application Diamond Park [74] in which users manipulate their avatar through pedaling a stationary bicycle. Others enable manipulation through handheld controllers, a computer mouse, or strictly hand tracking. A few prototypical applications offer little, if any, manipulation of one's virtual body.

Lastly, *avatar locomotion* refers to an avatar's ability to travel through virtual space. Walking is the most common design choice, although newer commercial applications and one prototypical application also support teleporting. In a few commercial applications and one prototypical application avatars can also fly, sometimes with the support of a plane, hovering boat, or other virtual vehicle. Social VR applications can also afford no avatar locomotion, meaning that users remain stationary in the virtual environment.

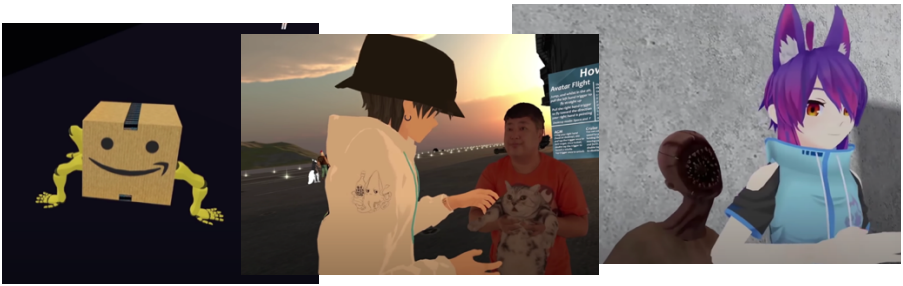

Fig. 1. Avatar customization options vary greatly in commercial applications such as VRChat. Left: avatar as cardboard box; middle: photorealistic avatar; right: avatar resembles a monster with a large scary mouth and no eyes. *Screenshots from "Stimulating my Friend's Phantom Touch in VRCHAT" by NoLogicDavid, licensed under CC BY https://www.youtube.com/watch?v=0Kx3Kpwzd6o*

## 4.2 Taxonomy Section 2: Interaction with Others

This design area involves users directly interacting with one another. Three categories of interaction features were discovered. *Communication types* refer to how users communicate, or the types of content that they can exchange. Physical expression is the most common example in the taxonomy. This utilizes avatar manipulation for nonverbal communication through gestures and bodily movements. Voice communication is also common through microphones typically attached to VR headsets. Visemes refers to lip or mouth movement to provide avatar lip-sync for voice communication. A less common communication type is transactions, referring to the transfer of currency or items between users. The rarest communication type in the taxonomy is bio-adaptive feedback, exemplified only in the prototypical application DYNECOM VR [58] with visualization of brain activity and respiration rate.

Fig. 1. In Rec Room an avatar's mouth moves to indicate that the user is engaging in voice communication ("Visemes" under Communication Types in the taxonomy). *Screenshots from "NOT VRCHAT: REC ROOM"*

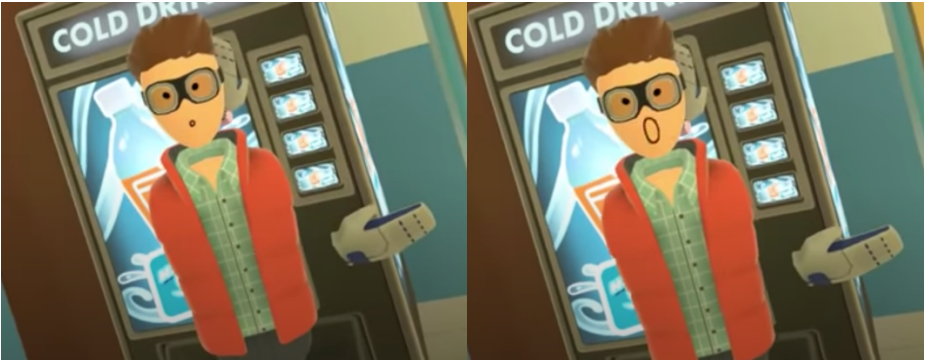

*by NadditionVR, licensed under CC BY https://www.youtube.com/watch?v=kZ4rN0nnmmA*

*Communication privileges* are ways that users can control who is able to communicate with them. Blocking and muting individual users from direct contact are common reactive features (they can be employed in response to harassment or unwelcome communication). Other approaches include contact lists and invitations to interact in private virtual areas. A less common design choice, discovered only in commercial applications, is space control. This refers to a user's ability to designate a space around their avatar that is inaccessible to other users or otherwise prevents users from interacting within that space. For example, in Rec Room avatars will gradually disappear once they enter a user's personal space.

Several social VR applications also *scaffold interaction with social activities*. Some of these activities revolve around collective observation or discussion. We label these as "events" and examples include public viewings of videos or streams on virtual theater screens (see AltSpaceVR [31,50–53,61]), public forums with the application development team (see Sansar [70]), and public participation in a theater performance (see Holojam in Wonderland [24]). We describe "recreation" activities as those that users actively engage in together, usually through ample avatar movement, such as dancing (see Virtual Dancing [65]), or minigames like paintball and charades (see Rec Room [50,51,53]).

A few applications were also found to support collaborative work. These examples include prototyping with virtual pen and paper (see PrototypingVR [35]), collaborative floor planning (see Embodied VR [63]), and collaborative review and modification of CAD models in General Motors' Tele-Immersive VisualEyes [39]. Other examples reflected a teacher/student dynamic with collaborative editing of virtual whiteboards (see the now-defunct Facebook Spaces [31,40,50,51]) and other learning artifacts like a virtual model of an ear for medical students (see Virtual Temporal Bone [39]).

### 4.3 Taxonomy Section 3: The Environment

This design area encapsulates design choices regarding the virtual space in which users can interact and manipulate their avatars. Virtual environments can vary based on *openness*. Public environments allow any users to access the virtual space and freely interact. Public environments can be populated by several concurrent users who may or may not know each other, and can quickly become overwhelming if too many users are present. By contrast, private environments are accessible only to predetermined users or by invitation (see Mozilla VR [34,50]). All prototypical applications are listed under private environments because users could not freely

.

access the virtual environment without deliberate action on behalf of the researchers or application developers.

*Spawning locations* refer to where users enter virtual space when they access the social VR application. Some applications spawn users directly into, or on the periphery of, a social location accessible by other users. Other applications spawn users into private rooms accessible only to the individual to help them orient to VR before entering a public space (see Virtual Dancing [65]). In some cases these private spaces are designed to replicate familiar personal spaces, such as a bedroom in Rec Room [50,51,53]. Users may traverse back to their private room after interacting in public spaces to make modifications to their avatar or VR settings. Lastly, users accessing a social VR application for the first time may spawn into a tutorial, a variation of the private space designed to teach users the basics of the application.

Several applications in the taxonomy enable users to *manipulate* their virtual private and public environments. Common examples involve creating and altering objects. This includes drawing artifacts in the virtual space, which users can then move around and apply to their avatars as if they were objects (see Facebook Spaces [31,40,50,51]). Several applications also allow users to move or modify existing objects in the virtual space that they did not personally create (see PrototypingVR [35]). In some applications users can also create entirely new virtual spaces and design them per their preferences (see Anyland [50]). Some commercial applications like VRchat, Rec Room, and Sansar offer a near endless array of manipulation possibilities for environment design that can result in rooms resembling physical world environments such as treehouses and parks, or distinctly fantastic environments like outer space landscapes. There are also several applications, particularly on the prototypical side, that afford no manipulation or alteration of the virtual environment.

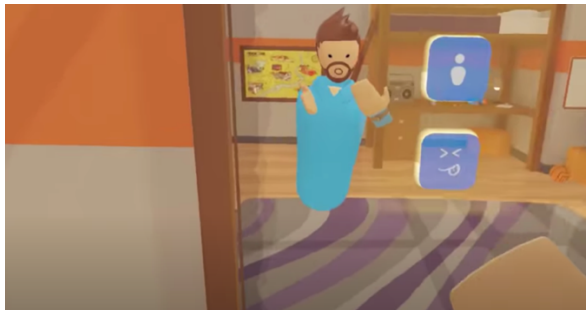

Fig. 1. Private spaces are sometimes designed only for individual users, and in the case of Rec Room can be designed to mimic physical world private places like a bedroom. Here users can modify their avatar and observe the changes (pictured), or simply relax before entering a social space. *Screenshot from "NOT VRCHAT: REC ROOM" by NadditionVR, licensed under CC BY*
*https://www.youtube.com/watch?v=kZ4rN0nnmmA*

## 5  LIMITATIONS

It is possible that some social VR applications were not discoverable through our chosen review methods, and some could have also been missed due to human error. These limitations also apply to design choices within the discovered social VR applications, particularly those that were not personally accessible to the researchers. It is also possible, if not likely, that the state of social VR design has changed since we conducted our respective reviews. New social VR applications could have been created, existing ones could have had their designs updated, and new literature about social VR indicative of novel design choices could have been published, which would not be represented in the taxonomy.

## 5  DISCUSSION AND CONCLUSION

Documenting social VR design choices is important for informing future social VR applications and ideating solutions to emerging user problems. Now more than ever, social VR is accessible to the masses because of affordable VR hardware like the Oculus Rift and HTC Vive as well as an array of commercially available VR applications. It can be easy to forget that social VR prototypes and private applications have existed for decades before Rifts and Vives entered our homes, and no documentation of social VR design can really be complete without their inclusion.

For this paper we constructed a taxonomy, or categorization, of social VR design choices informed by 44 social VR applications across commercial and non-commercial domains. This taxonomy extends efforts to document social VR design, which of recent have focused exclusively on commercially available social VR applications released in the last five or so years [36,50,51,66]. Aside from expanding the sheer number of applications incorporated into design documentation, the taxonomy visually organizes design choices in a way that is rapidly digestible to new researchers and designers. One anticipated use is as a design artifact in participatory design sessions. We intend the taxonomy to provide value to more experienced researchers as well through its capacity to be modified and grown. It could serve as an organizational tool to keep track of new features that are discovered or created as time progresses. Researchers can also use the taxonomy to map relevant literature to particular design choices, such as empirical studies of specific features (e.g., impact of avatar design choices on social presence), and identity features that are under-explored.

.

.